\documentclass[runningheads]{llncs}
\usepackage[T1]{fontenc}
\usepackage{cite}
\usepackage{amsmath,amssymb,amsfonts}
\usepackage{algorithmic}
\usepackage{graphicx}
\usepackage{subfigure}
\usepackage{textcomp}
\usepackage{xcolor}
\usepackage{comment}
\usepackage{hyperref}
\usepackage{float}
\usepackage{wrapfig}
\def\BibTeX{{\rm B\kern-.05em{\sc i\kern-.025em b}\kern-.08em
    T\kern-.1667em\lower.7ex\hbox{E}\kern-.125emX}}

\begin{document}

\title{Real-Time Cellist Postural Evaluation With On-Device Computer Vision}

\author{Paolo Wang\inst{1} \and 
Michael Zhang\inst{1} \and 
Shrinand Perumal\inst{1} \and 
Ekaterina Tszyao\inst{1} \and 
Luke Choi\inst{2} \and 
Kexin Sha\inst{1} \and
Felix Lu\inst{3} \and 
Paige Lorenz\inst{1} \and
Jackson P. Shields\inst{1} \and
Sivamurugan Velmurugan\inst{1} \and
Joshua Kamphuis\inst{1} \and
William P. Jiang\inst{1} \and
Gurtej Bagga\inst{1} \and 
Trevor Ju \inst{1} \and
Raymond Otis Kwon \inst{4} \and
Kristen Yeon-Ji Yun\inst{1} \and 
Yung-Hsiang Lu\inst{1}}

\institute{Purdue University, West Lafayette IN 47906 \\ \email{\{wang6601, zhan5202, speruma, etszyao, ksha, lorenzp, shield34, svelmuru, jpkamphu,
jiang868, gbagga, jut, yun98, yunglu\}@purdue.edu} \and 
Georgia Institute of Technology, Atlanta GA 30332 \\ \email{lchoi39@gatech.edu} \and University of Michigan, Ann Arbor MI 48109 \\ \email{felixlu@umich.edu}
\and Seoul International School, Seongnam, South Korea \\ \email{00rkwon00@gmail.com}}

\authorrunning{Wang et al.}
%
\maketitle

\begin{abstract}
    Posture is a critical factor for beginning instrumental learners. Most students receive instruction only once a week, and during the intervals between lessons they have little or no feedback on their physical posture. As a result, posture often deteriorates, increasing the risk of musculoskeletal injury and inefficient technique. Recent advances in computer vision and machine learning make it possible to evaluate posture without the constant presence of a human expert. However, current solutions have been extremely limited in availability and convenience due to their reliance on computationally expensive hardware or multi-sensor setups. We present Cello Evaluator, a real-time postural feedback system for practicing cellists. Through this optimization for on-device computer vision inference, we provide access to cellist postural evaluation to anyone with a current generation Android phone and thus reduces the postural feedback voids within individual practice. To validate our mobile application, we conduct a heuristic evaluation consisting of cellist and UX experts. Overall feedback from the evaluation found the app to be user friendly and helpful. 
    
\end{abstract} 

\begin{figure}[h]
    \centering
    \includegraphics[width=0.3\textwidth]{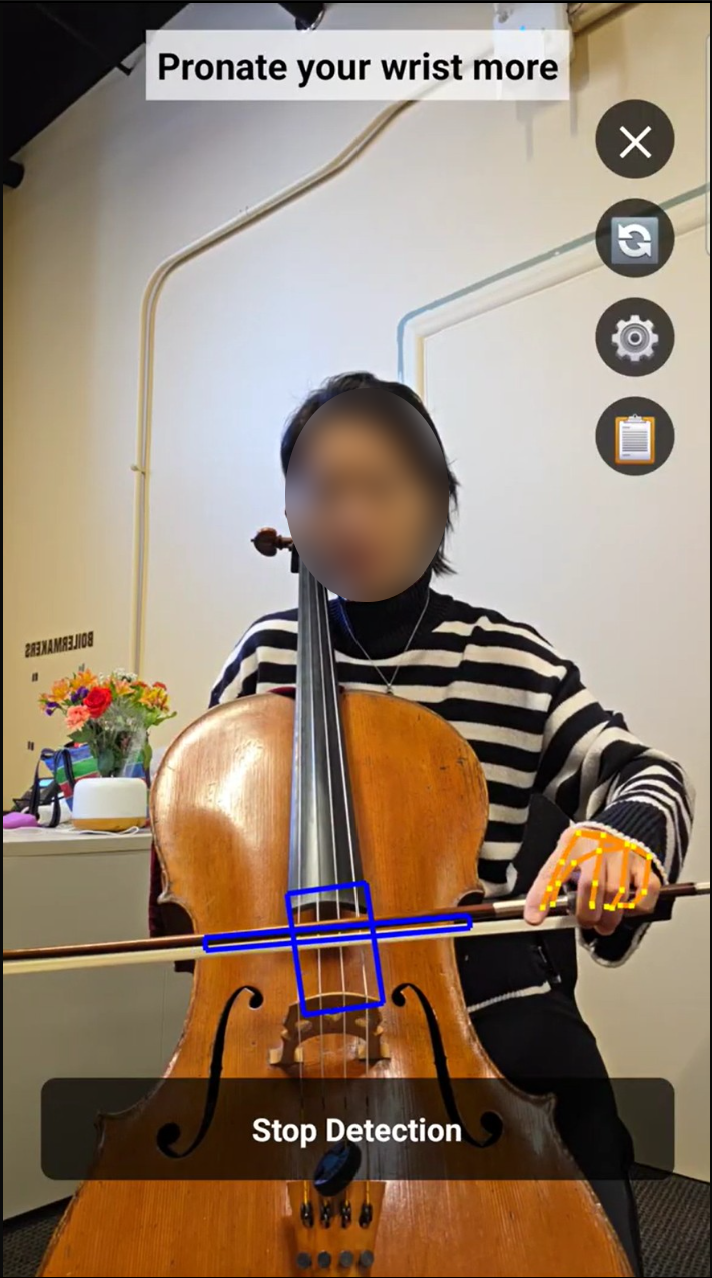}
    \caption{This is a screenshot of the Cello Evaluator being used during practice. The text at the top of the screen is an example of the instructions provided to the user for correcting poor posture. In addition, the bounding boxes of the relevant areas of the cello and the bow are depicted in blue, meaning that the posture in regards to the two objects are correct. The detected nodes of the user's bow hand is also detected and displayed; however, as the wrist posture is incorrect, the color used for the annotation is orange, signifying incorrect posture.
    }
    \label{fig:mobileapp}
\end{figure}

\section{Introduction}
Research indicates that 50\% to 76\% of professional musicians will experience musculoskeletal injuries in their careers, with harmful
postures being a primary contributing factor \cite{b1}. For many beginner cellists, maintaining ergonomically efficient and healthy postures can be challenging, as expert postural feedback is often limited to weekly lessons. Consequently, the majority of a cellist's practice occurs in large unsupervised intervals lacking any postural feedback, allowing for poor postural habits to form and become ingrained. This increases the risks of musculoskeletal injuries and can negatively impact musical development through harmful technique and sound quality.

\indent To address this gap in postural feedback for cellists, we introduce Cello Evaluator, a mobile application that provides real-time postural feedback through on-device computer vision. By incorporating postural feedback into the practice routines of cellists, Cello Evaluator reduces the risks of developing poor postural habits and, therefore, the overall health and development risks that cellists may encounter. Figure~\ref{fig:mobileapp} is a screenshot of the mobile app. When the front camera is used, the image is a mirror; thus, the cellist's right hand appears as the left hand in the app.

\indent Our system is therefore able to evaluate a practicing cellist's posture in near real-time, ultimately achieving a single frame inference latency of roughly 50 milliseconds. We evaluate the effectiveness and accuracy of our app by conducting a heuristic evaluation of two cello experts and two UX designers, in which we ask them various questions about the usability of our system. Results from the evaluation reported that the app was usable and user-friendly. 

\section{Background and Related Work}

\subsection{Posture within Music}

Posture has long been a topic of significant discussion within the field of music education. From a biological standpoint, repetitive motions with poor posture can result in the development of musculoskeletal disorders like tendinitis, nerve entrapment syndromes such as carpal tunnel, and dystonia \cite{b1, b7}. Such disorders can be extremely painful and will hinder a musician's development. In addition, poor cellist posture will negatively affect the cello's sound quality \cite{b6}. Thus, it becomes imperative to address postural issues as soon as possible in order to prevent the development of poor postural habits and musculoskeletal injuries.

\subsection{Real Time On-Device Computer Vision}

The emergence of lightweight neural network architectures (MobileNets, EfficientNet)\cite {b9, b8} and frameworks (Google Mediapipe, Tensorflow Lite)\cite{b2}, along with improvements in available computational power within mobile devices, have allowed for the development of various on-device applications of computer vision with near real-time feedback due to their minimal latency. On-device inference has multiple advantages over cloud inference: First, it enhances privacy by keeping all data local, ensuring it remains inaccessible to remote adversaries \cite{b10}. Furthermore, it avoids the network latency from cloud inference, and therefore reduces the overall inference latency and better allows for near real-time feedback. Finally, it makes the application more convenient by eliminating the need for an internet connection for inference.

\subsection{Postural Evaluation through Computer Vision}

Postural feedback through computer vision is a relatively new application within the domain of deep learning. There are three primary challenges to postural detection and evaluation using computer vision: 1) detecting the joints of the body accurately, 2) maintaining fast inference speeds, and 3) accurately evaluating the correctness of the posture detected. There are many frameworks for body detection and tracking; popular
computer vision models include Google Mediapipe, OpenPose, and You-Only-Look-Once (YOLO) Pose Estimation \cite{b2, b11, b4}. YOLO is the fastest of the three frameworks, while OpenPose and Mediapipe offer greater accuracy at the cost of increased computational complexity. 

\indent Prior studies involving instrument-related postural detection have been relatively successful, obtaining good detection and classification accuracies in piano \cite{b12}, violin, flute, and cello \cite{b5}. However, these algorithms all struggle from the same problem of requiring a high amount of computational power, which results in either a high inference latency or the requirement of high end GPUs for real-time feedback. Thus, these proposed solutions become inaccessible to the average cellist student with only access to their mobile devices. In addition, these systems typically used a postural detection model like Mediapipe or OpenPose and paired its outputs with classification algorithms like Convolutional Neural Networks (CNN), K-Nearest Neighbors (KNN), or Support Vector Machine (SVM) models for postural classification\cite{b12, b5}. Finally, prior work in cellist postural evaluation focuses purely on kinesiological postural errors, while the Cello Evaluator also analyzes bow placement and angle correctness.

\subsection{Our Contributions}
While prior attempts at postural evaluation using computer vision are largely inaccessible to the general public, we introduce a more convenient system capable of running real-time computer vision for cellist postural evaluation on most modern mobile phones. Our approach is directly optimized for on-device inference, meaning we are able to achieve low inference latencies on mobile devices fully offline, without the need for cloud inference. In addition, we show that Deep Neural Networks (DNNs) can be a viable alternative to other classification algorithms like CNNs, KNNs, and SVMs to maintain strong classification accuracy for postural correctness classification while also achieving fast inference speeds.

\section{On-Device Evaluation of Cellist Posture}

\begin{figure}[h]
    \centering
    \includegraphics[width=0.5\linewidth]{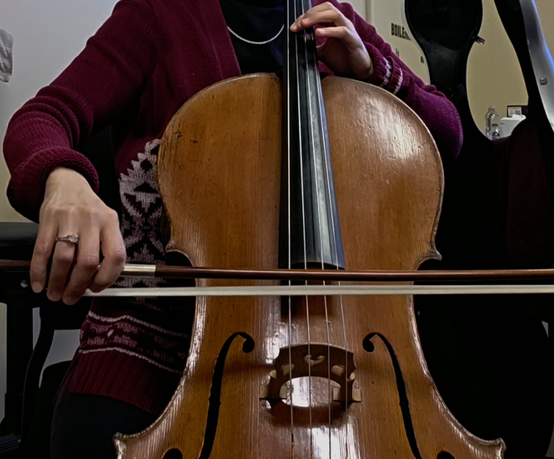}
    \caption{An example of correct cello posture with a correctly slightly pronated wrist, correct elbow height, and appropriate bow height (as it is touching the strings between the bottom end of the string board and the bridge of the cello) and angle (perpendicular to the cello strings). Additional details regarding how postural correctness is evaluated is discussed in sections \hyperref[sec:3.2]{3.2} and \hyperref[sec:3.3]{3.3}.}
    \label{figure:correctposture}
\end{figure}

Optimal cellist posture is defined by a neutral bow wrist with a slight inward hand tilt from natural pronation; a natural, relaxed elbow height; and a bow that intersects the cello strings perpendicularly within the region between the bottom edge of the fingerboard and the cello bridge.(Figure~\ref{figure:correctposture}). 

\subsection{System Design}

Our overall design consists of the following components: 
\begin{itemize}
    \item 
The phone's front or rear camera captures frames when a cellist is playing

\item The mobile phone 
processes the input visual
data to evaluates the user's bow
placement and body posture.    

\item Cello Evaluator
adopts two pre-trained frameworks: 
YOLOv11 and Google Mediapipe.
The former tracks the cello strings and the bow, and the latter tracks the cellist's hand and body. 
\end{itemize}


\indent We initially planned on performing cloud inference on a hosted server, as it was difficult to run multiple computer vision models concurrently on a mobile device due to the high computational cost. However, we ultimately opt for an on-device solution due to its advantages as discussed in the prior section. Therefore, we elect to restrict all inference to on-device only to reduce latency, protected user privacy, and increased overall app convenience.  To reduce
latency, the models are quantized. This results in a marginal degradation in model accuracy.


\indent The frontend and backend of the app are built with React Native and Kotlin native modules, respectively. React Native is used for its cross-platform flexibility, which provides convenience for further development on both Android and iOS systems.\\
\indent On Android devices, the CameraX library is used to capture video footage and display a live preview. We elect to use CameraX over the react-native-vision-camera library due to its lower latency and tight integration with the Android life-cycle, despite its lack of cross-platform compatibility.

\subsection{Wrist and Body Posture Classification}
\label{sec:3.2}

\begin{figure}[!htbp]
    \centering
    \subfigure[]{
    \includegraphics[height=.3\linewidth]{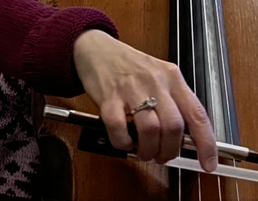}}
        \subfigure[]{   
    \includegraphics[height=.3\linewidth]{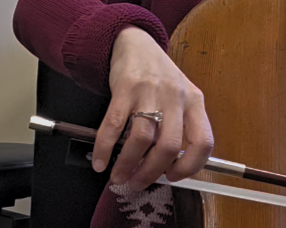}}
    \caption{(a) Example of an incorrect, supinated wrist posture, in which the forearm and hand are rotated such that the palm faces upward. (b) Example of correct wrist posture, with slight pronation, where the forearm and hand are rotated such that the palm faces downward.}
    \label{fig:comparewrist}
\end{figure}

This app targets the postures as 
the orientation of the wrist and height of the elbow. A common error in wrist posture is supination (Figure~\ref{fig:comparewrist}). Supination occurs when the forearm is outwardly rotated, which causes the palm to face away from the cello and slightly upwards. Conversely, pronation occurs when the forearm is inwardly rotated so that the palm faces towards the cello body or the floor.
A slight degree of pronation is widely considered correct posture, but too much pronation is also considered poor posture. For elbow height, a proper posture must not be too high or too low and must lie at a reasonable height. Specifically, when the bow hand is extended away from the cello strings, the elbow should be positioned close to the side of the cello body.

To evaluate the posture of the wrist and elbow, we use two pre-trained Google Mediapipe models, Pose Detection \cite{b2} and Hand Detection \cite{b3}, to extract the coordinates of the cellist’s body and hand nodes at each frame. The Mediapipe framework was chosen for its lightweight, on-device optimization, allowing for faster inference on a standard smartphone device compared to other Human Pose Estimation models. Once the coordinates are extracted, two small Deep Neural Networks (DNNs) are then used to produce final posture classification results for wrist and elbow posture, respectively. The size of each DNN stands at ~1100 parameters for wrist posture and roughly 440 parameters for elbow posture. Due to the complexity of the extracted nodes, we elected to use DNNs over purely mathematical approaches for classification. Furthermore, DNNs were chosen over Convolutional Neural Networks (CNNs), as the spatial complexity of data was not high enough to warrant the use of CNNs and a DNN consists of fewer parameters allowing for faster real-time inference.

\begin{figure}[h]
    \centering
    \includegraphics[width=0.5\linewidth]{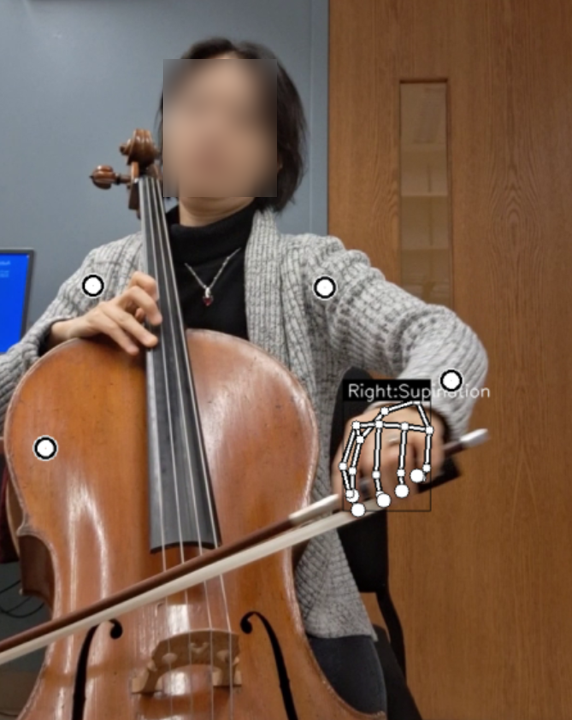}
    \caption{An example image of training data collection, where the cellist is purposely playing with an incorrect supinated wrist posture to allow for collection of supinated wrist node coordinates.}
    \label{figure:datacollectionex}
\end{figure}

For both DNNs, roughly 10,000 frames of labeled footage collected from volunteer cellists were recruited to produce the training data. These cellists recorded videos of themselves playing with either correct or intentionally incorrect posture, allowing us to collect data for each posture class (Figure~\ref{figure:datacollectionex}). OpenCV \cite{b23} was used to extract, pre-process, and annotate individual frames from these videos prior to landmark extraction and model training. It should be noted that all training videos were recorded from a similar position relative to the cellist, matching the expected use case of our application. However, this method of data collection limits the accuracy of our application when used from different angles or positions.  This training process achieves validation accuracy of over 95\%. 

For wrist posture classification, all 21 hand node coordinates generated from Google Mediapipe Hand Detection are used as inputs for the respective DNN. Specifically, the coordinates used are the normalized X and Y coordinates of each hand node, with the origin designated as the palm node. We do not consider the depth of each hand node due to its unreliable accuracy, especially during the fast movements of the hand when playing cello. Finally, given the coordinates, the DNN can output three possible posture classes: normal (correct), supination, and over-pronation. 

To achieve elbow posture classification, three nodes are extracted from Google Mediapipe Pose Detection: the bow-side shoulder, elbow, and wrist. From these three nodes, linear algebra operations are applied to their 3D coordinates to derive nine features used as inputs to the DNN. These features include the elbow joint angle, the Euclidean distance from the shoulder to the elbow, the Euclidean distance from the elbow to the wrist, and the X, Y, and Z components of the normalized direction vectors from the shoulder to the elbow and from the wrist to the elbow. It is worth noting that the bow-side arm does not move as rapidly as the bow-side hand, resulting in a higher depth perception accuracy from Mediapipe, which allows us to use 3D coordinates as opposed to 2D coordinates. Additionally, using these values as opposed to raw coordinates resulted in a higher model accuracy, as previous iterations of the DNN trained using only raw coordinates were not capable of correctly classifying elbow posture. The three possible posture classes produced are normal elbow, elbow too low, and elbow too high.

\subsection{Bow and String Classification}
\label{sec:3.3}
\begin{figure}[h]
    \centering
    \includegraphics[width=.5\textwidth]{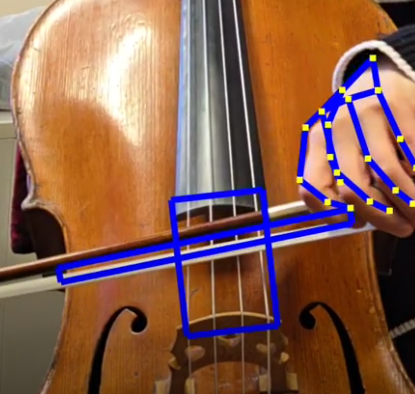}
    \caption{The bounding boxes of the cello bow and the cello string regions are used to identify the bow's height and angle relative to the strings, which are then used for postural classification.}
    \label{fig:yolo_training}
\end{figure}

To evaluate bow placement and angle, we focused on two regions of interest: the bow and the strings. As the far end of the cello bow is not always within the camera frame, we track only the portion of the bow that overlaps with the cello (Figure~\ref{fig:yolo_training}). This approach ensures that the bow-string contact area remains visible at all times. Additionally, we choose to track the region of the cello strings between the bridge and the bottom edge of the fingerboard, as this is the conventional playing zone for beginners. This approach allows us to determine whether the bow is outside the correct playing zone.
\\
\indent
Given these regions of interest and application constraints, the selected model needed to track bow and string positions precisely for accurate classification while maintaining low inference latency to achieve real-time performance. We evaluated several models, including Segment Anything Model 3 (SAM 3)\cite{b16}, Detectron2 \cite{b17}, You Only Look Once (YOLO) \cite{b18} , and Real-Time Detection Transformer (RT-DETR) \cite{b19}. Most models failed to meet our criteria, with the exception of YOLO. We found that the YOLO's single-stage architecture provides superior real-time inference compared to other computer vision frameworks. Additionally, the model offers various versions tailored to specific computer vision tasks, including object detection, oriented bounding box (OBB) detection, instance segmentation, and keypoint detection, allowing our team to experiment and identify the optimal approach to tracking the bow and strings.

\begin{figure}[!htbp]
    \centering
    \subfigure[]{
    \includegraphics[height=.3\linewidth]{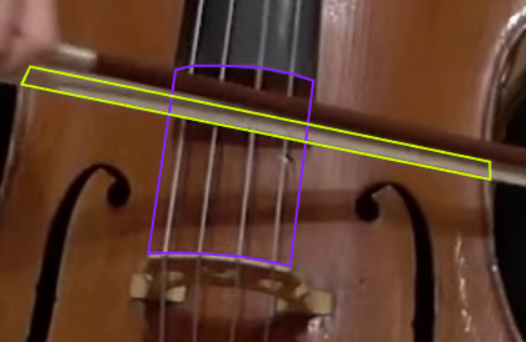}}
        \subfigure[]{   
    \includegraphics[height=.3\linewidth]{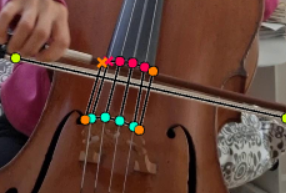}}
    \caption{Two label techniques were used to label each image needed to train the various YOLO models that utilize different computer vision techniques. (a) A labeled image used for training YOLO for instance segmentation, object detection, and orienting bounding boxes. (b) A labeled image used for training YOLO for keypoint detections, where points marked with "x" are occluded points.}
    \label{fig:bow-label}
\end{figure}

\indent
The YOLO models were trained using a custom image dataset curated and created by our team, consisting of more than 4,000 manually labeled images (Figure~\ref{fig:bow-label}) annotated through Roboflow. These images contain musicians who provided their prior consent playing the cello with varying bow and string postures. To improve model performance in diverse environments and usability, we applied several preprocessing augmentations during training. Each training example was augmented to produce three output variations with the following transformations: random rotation between ±8°, horizontal and vertical shear of ±10°, brightness adjustments between ±15\%, exposure variations between ±10\%, blur up to 2.5 pixels, and random noise affecting up to 0.1\% of pixels.

\begin{figure}[htbp]
    \centering
    \subfigure[]{
    \includegraphics[height=0.17\linewidth]{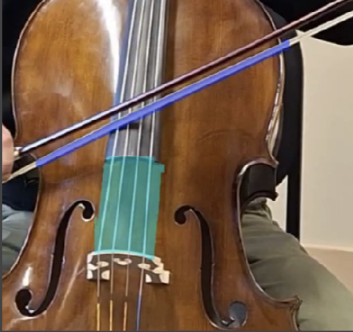}}
    \subfigure[]
    {\includegraphics[height=0.17\linewidth]{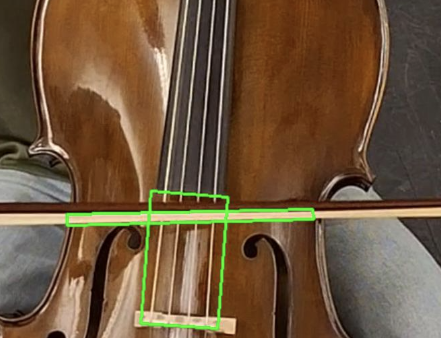}}
    \subfigure[]
    {\includegraphics[height=0.17\linewidth]{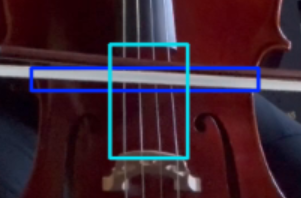}}
    \subfigure[]{\includegraphics[height=0.17\linewidth]{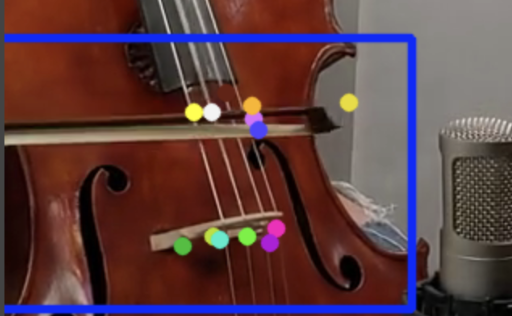}}
    \caption{These images showcase YOLO model predictions after training the model with our custom, manually labeled dataset. (a) YOLO instance segmentation detection. (b) oriented bounding box detection. (c) YOLO object detection. (d) YOLO keypoint detection.}
    \label{fig:compare-result}
\end{figure}

\begin{figure}[H]
    \centering
\includegraphics[width=.7\linewidth]{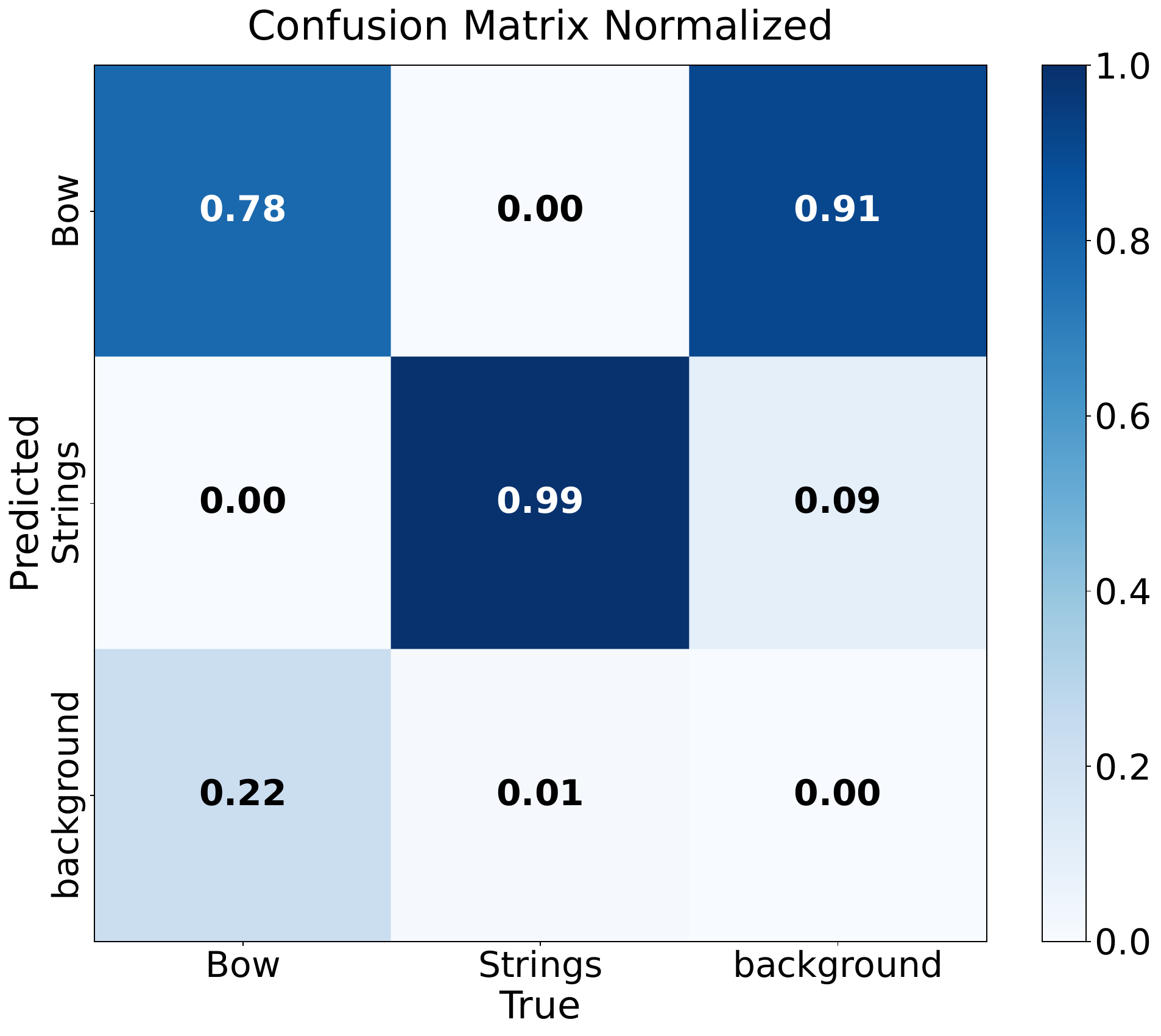}
    \caption{The normalized confusion matrix above illustrates how well YOLO-OBB performed at distinguishing the three classes: Bow, String, and background. The model performs excellently at identifying strings (99\% correct) and reasonably well at detecting background elements (91\% correct). The model struggles to consistently identify the bow (78\% correct) by misclassifying the bow as the background (22\% incorrect). }
    \label{fig:confusion-matrix}
\end{figure}

\indent After training various YOLO models, we compared and analyzed them to determine the most suitable computer vision approach (Figure~\ref{fig:compare-result}). We found that instance segmentation accurately modeled the curvature of the strings but was computationally expensive as the model calculates the entire area of the strings and bow, which exceeds our requirements. Keypoint detection did not provide accurate results due to insufficient data set size and robustness. Both object detection and OBB demonstrated the most consistent and best-performing results, but OBB accounted for the bow angle with tighter-fitting boxes that would provide more precision when classifying posture.  Based on these findings, we selected OBB for our classification task, specifically YOLOv11 \cite{b21}, which offered better precision and performance compared to YOLOv8 \cite{b20} and better performance than YOLOv12 \cite{b22} relative to the marginal accuracy gains achieved. The results shown (Figure~\ref{fig:confusion-matrix}) illustrate the performance of the YOLO-OBB model after optimizing the hyperparameters during training.
To classify bow height and angle placement, we developed a mathematical approach using the string box and bow box coordinates inferred by YOLO-OBB to determine appropriate feedback. Based on expert consultation, we determined that playing the cello with the bow positioned close to the bottom edge of the fingerboard in a back-and-forth motion perpendicular to the strings produces the clearest tone and sound. Initially, we verify whether the bow box intersects with the string box to confirm that the bow is within the string zone before classifying height and angle. Subsequently, we use the coordinates to compare the bow's height to the string zone height to classify the bow as too high, too low, or properly positioned. Finally, using the intersection of the bow box with the string box, we classify whether the bow angle is perpendicular to the strings. This constitutes our complete bow and string classification system for the Evaluator app. 

One major challenge for bow and string classification that hindered the app's effectiveness was the lack of real-time performance when running our YOLO model on Android devices. Using a Samsung S25 test phone, our models achieved 140 milliseconds per frame (mspf) and 90 mspf on CPU and GPU,  respectively. To reduce latency, the models were optimized using Qualcomm AI Hub through quantization, graph optimization, and hardware-specific compilation for Qualcomm processors. As a result, YOLO achieved 110 mspf and 40 mspf on CPU and GPU, respectively. Furthermore, by leveraging the NPU equipped on the Qualcomm Snapdragon chips in the S25 test phones, our model achieved 20 mspf. Through these hardware and software optimizations, the bow and string classification operates in near real-time, enabling efficient feedback for app users.

\subsection{User Interface}

\begin{figure} [h]
    \centering
    \includegraphics[width=0.5\linewidth]{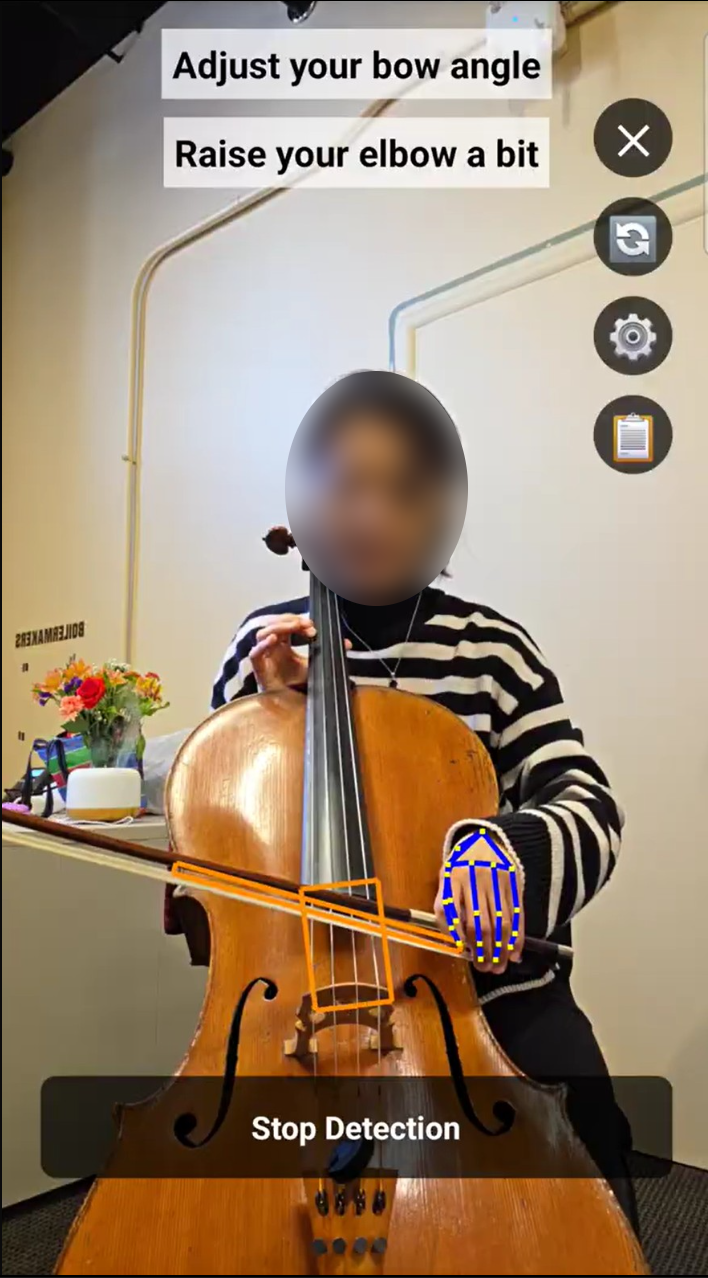}
    \caption{An example of hand and pose annotations with posture correction instruction at the top displayed by the app. Buttons on the right represent (from top to bottom): close camera, flip camera, open settings, and view session history.}
    \label{fig:fig8}
\end{figure}

The Cello Evaluator supports two modes of operation: (1) processing pre-recorded videos and (2) providing real-time feedback using the phone’s camera. The latter is the expected primary use case and the main focus of this paper. In both cases, the following annotations are directly overlaid onto the video feed: (1) Bounding boxes for the bow and string area, (2) Nodes of the tracked bow hand, and (3) Textual instructions for posture correction, as shown in Figure~\ref{fig:fig8}.

Boxes around the bow and strings directly provide immediate feedback on detection quality. They help users adjust their camera’s distance and angle to ensure the bow and the playing area of the strings are always fully detected and tracked. Annotations of the user’s bow hand are kept for the same reason. Because the full body skeleton frequently overlaps with other annotations and introduces distracting visual clutter, we choose not to display it on the screen.
\index Visual annotations dynamically change colors based on their respective postural correctness: blue when correct and orange when incorrect. This color contrast was selected to clearly differentiate the correctness states, making feedback easily noticeable during practice. Furthermore, we choose to use blue and orange instead of conventional green and red to improve accessibility for potential users with color-blindness.

\begin{figure}[H]
  \centering
  \subfigure[]{
    \includegraphics[width=0.45\linewidth]{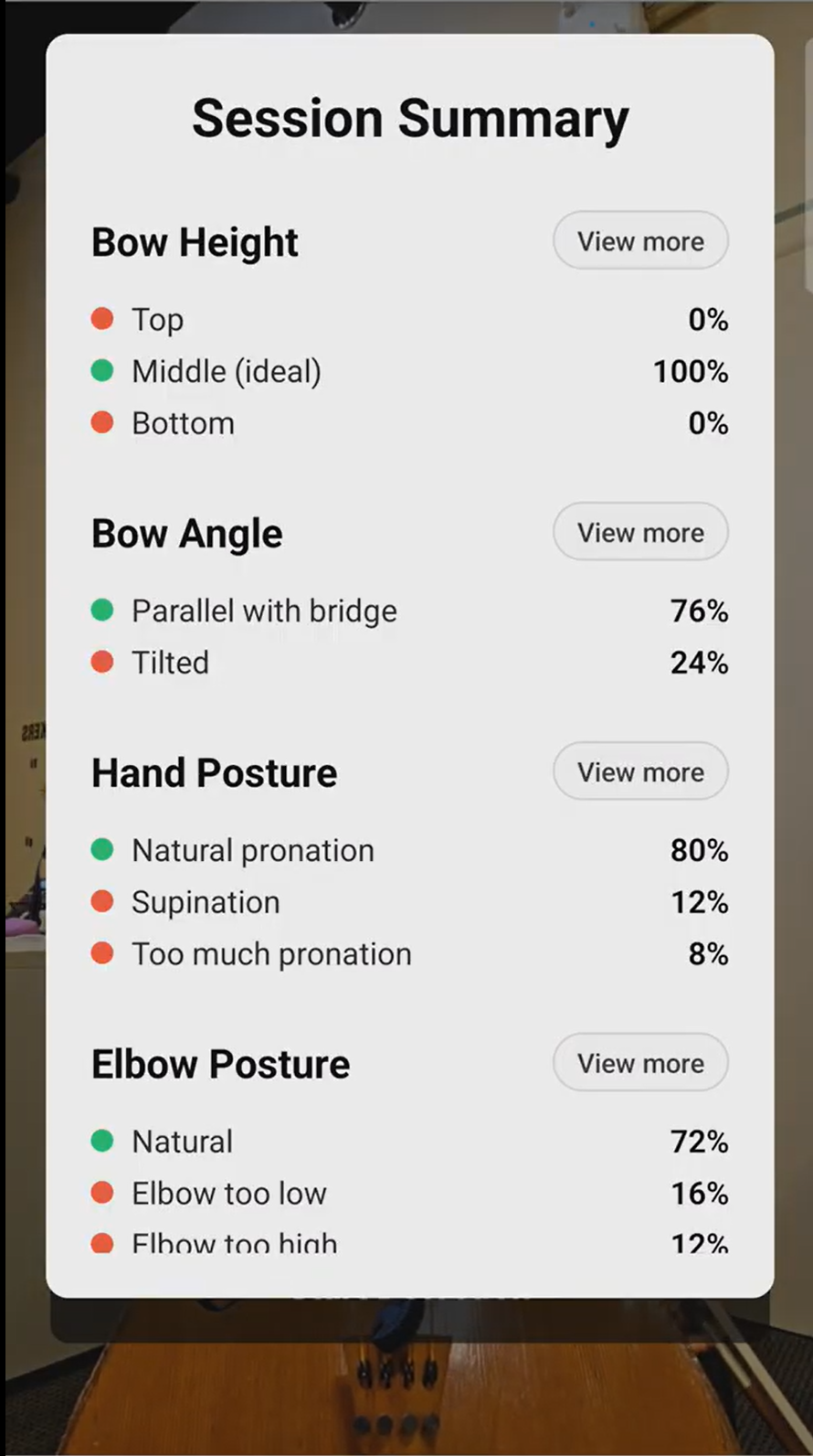}
    \label{fig:summary-a}}
  \hfill
  \subfigure[]{
    \includegraphics[width=0.45\linewidth]{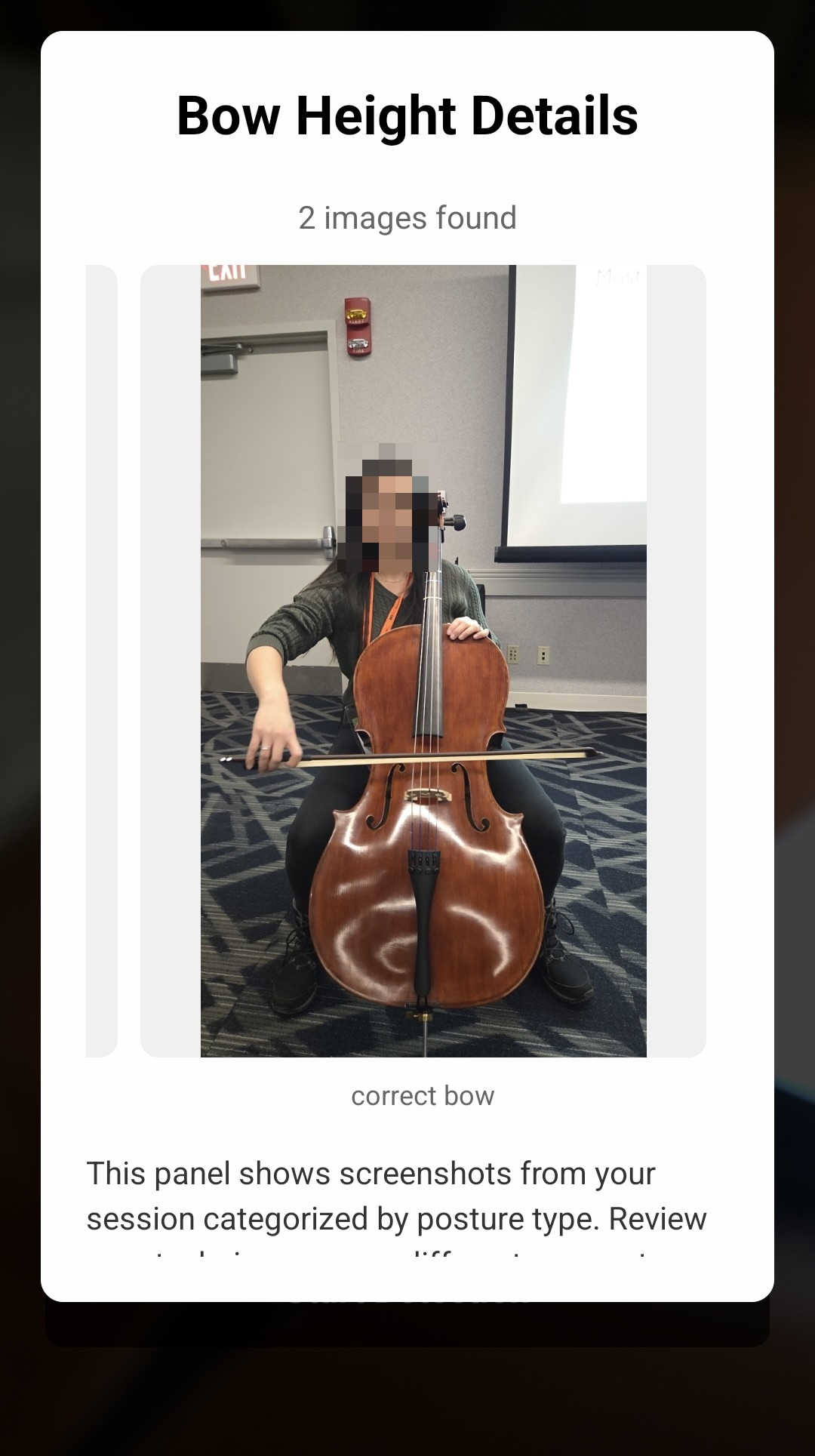}
    \label{fig:summary-b}}
  \caption{An example of the session summary with a detailed breakdown of (a) postural occurrence percentages and (b) representative screenshots for all categories under bow placement and cellist posture.}
  \label{fig:summary}
\end{figure}

Textual correction instructions are displayed at the top of the screen to support real-time posture adjustment. Instructions are ranked by error frequency, with the most frequent errors shown first, and are limited to at most two concurrent instructions to reduce redundancy and maintain readability. Furthermore, to avoid unnecessary or flickering feedback, an instruction appears only if the corresponding error persists for at least 5 seconds, and remains on the screen for at least 3 seconds until the error is corrected. This design ensures that users have sufficient time to read and respond to the feedback without being overwhelmed by rapid changes.

After each practice session, users receive a session summary report with a detailed breakdown of error rates to guide future improvements. A session summary contains occurrence percentages and representative screenshots for all categories under bow height, bow angle, hand posture, and elbow posture (Fig. ~\ref{fig:summary}). Throughout each practice session, the application tracks the number of frames classified under each category and computes percentages by dividing them by the total number of frames. Postural percentages within each section are then normalized to improve clarity. While quantitative metrics reflect the areas where the user performed well and the areas that require improvement, the screenshots help users visually distinguish correct and incorrect techniques. Users may also review past sessions through the session history feature to observe performance trends over time.

The application supports flexibility and customization with respect to the user’s skill level. Multiple users can share the same device using different logins without interfering with one another, which also simplifies the setup for the initial stages of user testing. Users may also adjust the acceptable range of bow angle deviating from perpendicular alignment based on their skill level and personal preference. 

\section{Multidisciplinary Expert Heuristic Evaluation}

\indent To identify early usability issues in the current prototype and to inform the next steps for improvement, we conducted a heuristic evaluation\cite{b13}. Heuristic evaluation is widely used to assess interface usability and interaction quality through expert review. We adapted the method to capture both quantitative severity ratings and qualitative insights. At this stage of development, heuristic evaluation is an economical and practical choice because it enables rapid diagnosis of interface and usability problems and generates actionable feedback that can inform the next stage iteration with fewer participants and less time, before investing in large-scale user testing.

\indent We recruited multidisciplinary experts, including two interaction designers familiar with user experience design and heuristic evaluation, and two expert cellists and cello educators, each with over thirty years of teaching experience. This expert composition was selected because the system’s effectiveness depends not only on general usability but also on whether the feedback and interaction model align with established cello technique and teaching practice. Participant recruitment and all evaluation procedures complied with IRB-approved protocols (IRB-2023-551).

\subsection{Heuristics Construction}
\indent The heuristics used in this evaluation were constructed based on Nielsen’s ten usability heuristics\cite{b14} and the Guidelines for Human-AI Interaction\cite{b15}. Nielsen’s heuristics provide a foundational framework for assessing interface usability and interaction quality, while the Human-AI Interaction guidelines focus on key considerations for AI-enabled systems.

\indent From these two sources, we selected heuristics that were most relevant to the prototype’s functionality and current development stage. We also included sub-heuristics related to music pedagogy to evaluate the prototype’s perceived usefulness for cello practice. The resulting twelve sub-heuristics are listed below. To reduce ambiguity in interpretation and improve consistency across evaluators, each sub-heuristic was paired with a guiding question for review:

\begin{description}
  \item{C1. Make clear what the system can do}: Is it clear what the app function is?
  \item{C2. Action and icon clarity}: Are key actions and icons clear, recognizable, and easy to understand without guessing?
  \item{G1. Help and Documentation}: When users are unsure (setup, metric meaning, or next step), does the app provide enough guidance at the right moment?
  \item{S1. Visibility of System Status}: During recording/detection, is status always obvious (active/paused/not tracking)?
  \item{S2. Make clear why the system did what it did}: Does the app explain ``why'' a posture was flagged (e.g., visual cues, reference lines, thresholds) in a way that supports learning and action?
  \item{S3. Make clear how well the system can do what it can do}: Does the app indicate confidence/limitations and warn users when judgments may be unreliable?
  \item{F1. Efficiency of Use}: Is it fast to start a practice session, especially for repeated use?
  \item{E1. Error Prevention}: Does the app help users avoid common errors, such as clicking the wrong button or using the wrong camera angle?
  \item{E2. Support efficient correction}: If the app mislabels a posture, can the user quickly correct or override it (or flag it)?
  \item{P1. Skill-level variability}: Does the app feel usable for different skill levels (beginner vs.\ advanced) without being too strict or too vague?
  \item{P2. Actionability}: Does the feedback feel reasonable for a cello learner to adjust posture without a teacher's presence?
  \item{P3. Perceived helpfulness for posture improvement}: Does the app feel helpful for improving posture in real practice?
\end{description}

\indent 
To summarize results at a higher level while retaining these specific usability or pedagogy concerns, we consolidated the twelve sub-heuristics into six higher level heuristics: Clarity (C), Guidance (G), System Transparency (S), Efficiency (F), Error Handling \& Recovery (E), and Musical Pedagogy Value (P).

\subsection{Data Collection}
Evaluation data was collected using a structured evaluation worksheet during one-on-one sessions with each expert. The worksheet was designed to support consistent assessment across evaluators and included all sub-heuristics, corresponding guiding questions, severity ratings, and open-ended feedback fields.
\indent For each sub-heuristic, experts assigned a severity rating on a 0-4 scale, where higher values indicated more severe issues. These ratings were used to obtain quantitative data reflecting the severity of issues. In addition, experts were asked to provide qualitative feedback, including perceived strengths (pros), limitations (cons), and recommendations, to capture insights that could not be fully represented by numerical ratings alone. 

\indent Prior to the evaluation, each expert was introduced to the purpose and intended use of the prototype. A short demo video was shown to illustrate a usage scenario and the core interaction flow of the app. The interfaces were presented as consisting of three stages: Setup, Play/Detect, and Session Summary. During the evaluation, experts were allowed to revisit the demo video or interface screens at any time to support their assessment. All evaluations were conducted independently without discussions among experts.

\subsection{Quantitative Insights}
\begin{figure}[H]
  \centering
  \includegraphics[width=\linewidth]{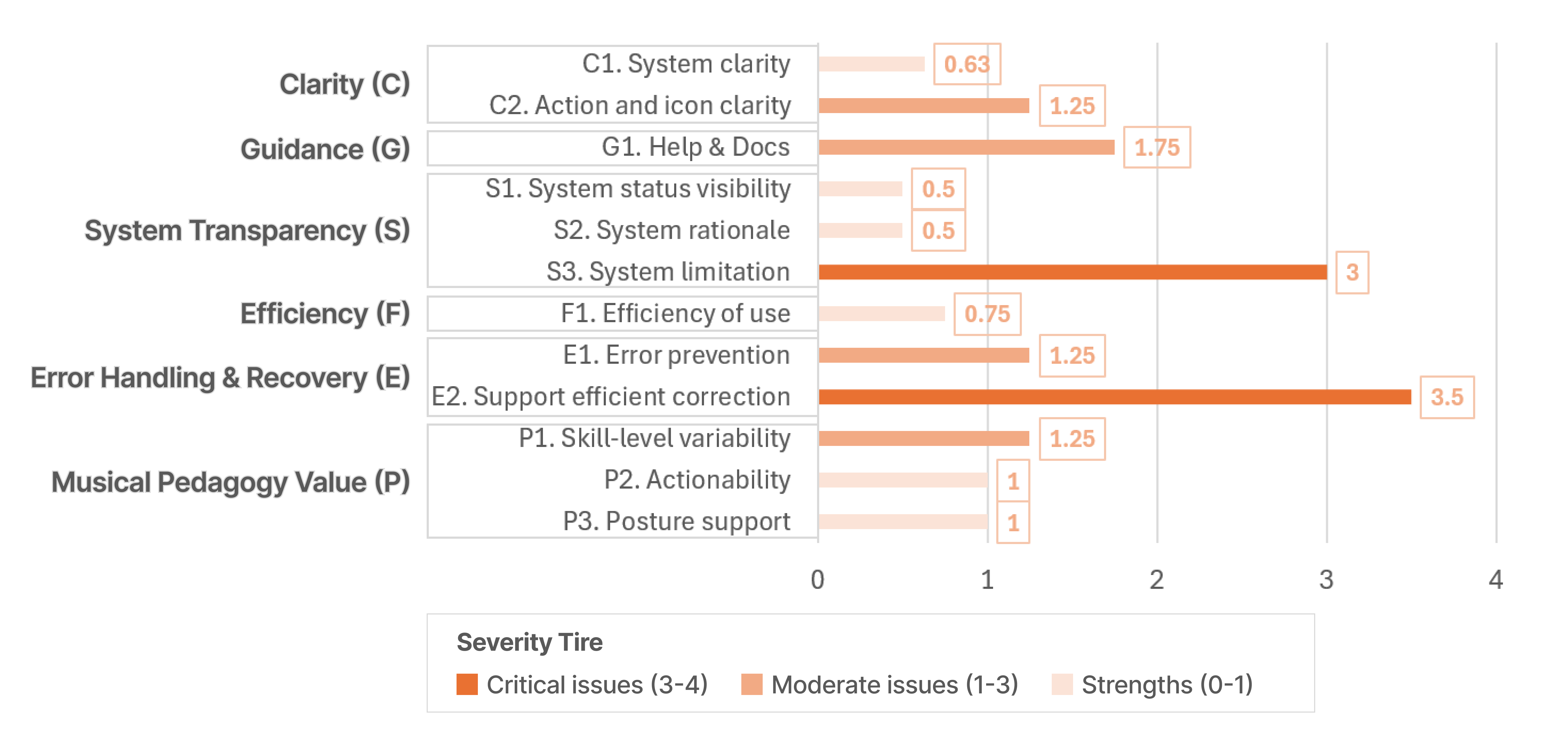}
  \caption{Mean severity ratings (0-4) for the 12 sub-heuristics, grouped into six higher-level heuristic categories. Bar colors indicate severity tiers: strengths (0-1), moderate issues (1-3), and critical issues (3-4).}
  \label{fig:Severity rating}
\end{figure}
Figure ~\ref{fig:Severity rating} presents the mean severity ratings across all sub-heuristics using a horizontal bar chart, grouped into six higher-level categories. As shown in the figure, the severity tiers are used to summarize the distribution of problems and indicate relative improvement priorities.

The most severe issues are concentrated on System Transparency and Error Handling \& Recovery. Specifically, S3 (Make clear how well the system can do what it can do) and E2 (Support efficient correction) received the highest average severity ratings, indicating critical gaps in communicating system limitations and supporting user report when misclassifications occur as an AI supported system.

A set of moderate issues was identified across multiple heuristics, including C2 (Action and icon clarity), G1 (Help and Documentation), E1 (Error Prevention), and P1 (Skill-level variability). These results suggest that while core interactions are generally functional, additional clarification, guidance, and adaptability are needed to better support diverse users and usage contexts.

In contrast, several sub-heuristics consistently appeared in the strength tier, including C1 (Make clear what the system can do), S1 (Visibility of System Status), S2 (Make clear why the system did what it did), F1 (Efficiency of Use), P2 (Actionability), and P3 (Perceived helpfulness for posture improvement). These findings indicate relative strengths in communicating system purpose, maintaining real-time interaction awareness, supporting efficient practice flow, and providing feedback that users perceive as actionable and helpful.

Overall, the quantitative results highlight clear priorities for design improvement while also confirming several stable strengths in the current interaction design.

\subsection{Qualitative Insights}
While the quantitative analysis highlights the relative severity of issues across heuristics, qualitative feedback provides richer insights into why these issues matter in practice and how experts interpret them from different professional perspectives. Qualitative comments were examined across experts within each sub-heuristic to identify common themes as well as discipline-specific perspectives. Based on this synthesis, we summarized higher-level design opportunities across three dimensions: usability and interaction design, Human-AI interaction, and cello pedagogy.

\subsubsection{Usability and Interaction Design Opportunities.}

Overall, experts viewed the prototype as usable for practice: the workflow is lightweight once detection begins, and the smooth real-time detection and feedback does not interrupt practice flow. These strengths are consistent with low severity ratings for efficiency of use (F1) and system status visibility (S1), indicating that the core practice flow can remain smooth during active use.

The primary usability friction was concentrated in the setup stage, where users may need repeated positioning adjustments while holding the instrument and may have to stand up to advance the flow (e.g., tapping “Ready” and “Start Detection”). This issue, together with moderate ratings for guidance (G1) and action/icon clarity (C2), suggests that the main challenges are not missing functionality, but the sequencing, discoverability, and timing of information and controls.

Experts therefore recommended reducing setup burden through clearer setup cues and fewer manual steps, strengthening first-use and just-in-time guidance (what the system can evaluate and what each metric means), and improving the clarity of key action icons and feature discoverability. They also noted that session summaries could better support learning if they moved beyond static screenshots toward more instructive representations of how to correct posture over time.

\subsubsection*{Human-AI Interaction Considerations.}

From a Human-AI interaction perspective, experts valued the app’s attempt to provide objective, real-time posture feedback and found the overall feedback loop promising for supporting practice. During detection, the interface presents visual evidence, such as hand overlays and bounding boxes for bow position and angle, which helped experts interpret some judgments in context and reflect relatively stronger performance on communicating why feedback was triggered (S2).

At the same time, experts emphasized that appropriate reliance on AI feedback requires clearer communication of system limitations (S3), particularly for beginners who may otherwise accept feedback uncritically or become overly dependent on the system. Experts suggested that the app should more explicitly indicate what the AI can and cannot assess well, and under what conditions judgments may be less reliable.

Experts also highlighted the value of allowing users to flag and label questionable judgments (E2). Providing a lightweight mechanism to mark potential misclassifications can increase perceived credibility by acknowledging uncertainty, while also creating an input channel that could support future model refinement and accuracy improvement.

\subsubsection*{Cello Pedagogy and Learning-Oriented Insights.}

From a pedagogical perspective, cellist experts generally agreed that the app is most suitable for beginner to intermediate learners who already have a basic understanding of cello posture. For these users, the system functions effectively as a practice support tool, providing timely feedback during independent practice to reinforce or correct fundamental posture habits. The immediacy of feedback was viewed as particularly helpful, allowing users to make quick adjustments while playing (P3).

Experts also noted that the session summary supports post-practice reflection and can be valuable for identifying recurring posture issues. With more detailed explanations, this feature could further help less experienced players understand why certain adjustments are recommended and how to improve over time.

At the same time, experts emphasized that posture expectations become more flexible at higher levels. Advanced players may adopt posture variations associated with different schools of playing or specific musical expressions, making rigid or universal judgments less appropriate. As a result, the app’s feedback was seen as less applicable to advanced cellists who require greater interpretive freedom (P1).

Across skill levels, experts highlighted the importance of grounding feedback in pedagogically meaningful language. They suggested prioritizing universally problematic posture issues, such as excessive shoulder tension or improper head and neck posture, while avoiding over-prescription in areas that allow individual variation. Several experts emphasized that these fundamental posture principles should be clearly communicated during the initial setup phase, reinforcing that correct setup is a critical first step before practice begins.

Overall, experts viewed pedagogical alignment as an opportunity for future design, suggesting clearer framing of the app as a supportive practice aid, more educational explanations in session summaries, and feedback strategies that balance universal posture principles with flexibility for individual and expressive differences.

\section{Discussion and Future Plans}
As described in the analysis of the heuristic evaluation, there are a few areas in which our app can be made more user friendly. Namely, usability could be improved by accounting for false-positive cases when a classification is faulty. To address this issue, we plan on implementing the option to disable certain classifications depending on user need or priority. In addition, training data diversity could be increased to improve classification and detection accuracies from uncommon or unconventional camera angles. Follow-up studies could investigate whether there are more efficient classification algorithms for evaluating postural correctness in different angles. 

\indent Additionally, an upcoming long-term user study is planned for February 2026 and will involve over 60 cellists of varying skill levels. Each participant will complete a usability study and engage in multiple scheduled practice sessions using the application. By collecting postural error percentages across sessions, we expect to observe a decrease in postural error rates as the number of practice sessions increases.
The study has been approved by the Purdue University Institutional Review Board (IRB-2023-551). Written informed consent is obtained from all participants.

\section{Conclusion}
In this work, we provide a novel solution to improving musician health that prioritizes user privacy, reduced latency, and overall availability through live, on-device computer vision. We show that pairing a DNN with YOLO and Google Mediapipe is both efficient and accurate at detecting and classifying errors within cellist posture. We also find on-device inference to be the most effective approach to minimize frame processing latency, and ultimately achieve 50 ms single frame inference time on a Samsung S25 with NPU acceleration. A heuristic evaluation also found that our app was user friendly and helpful, with the only area for improvement being false-positive cases. These results confirm that on-device computer vision is an effective and approachable method for evaluating cellist posture.

\section*{Acknowledgments}
This project is an ongoing effort under Purdue's AI for Musicans (AIM) organization. We are grateful for the contributions from: Preston Tang Mo, Kayoon Koh, Mukund Rao, Zihao Ye, and Hanako Keney.
This project is supported in part by the National Science Foundation IIS-2326198. 
Any opinions, findings, and conclusions or recommendations expressed in this material are those of the authors and do not necessarily reflect the views of the National Science Foundation.

\bibliographystyle{splncs04}
\bibliography{bibliography}

@article{b1,
  author    = {Heinan, M.},
  title     = {A review of the unique injuries sustained by musicians},
  journal   = {JAAPA: official journal of the American Academy of Physician Assistants},
  volume    = {21},
  number    = {4},
  year      = {2008},
  doi       = {10.1097/01720610-200804000-00015}
}

@misc{b2,
      title={MediaPipe: A Framework for Building Perception Pipelines}, 
      author={Camillo Lugaresi and Jiuqiang Tang and Hadon Nash and Chris McClanahan and Esha Uboweja and Michael Hays and Fan Zhang and Chuo-Ling Chang and Ming Guang Yong and Juhyun Lee and Wan-Teh Chang and Wei Hua and Manfred Georg and Matthias Grundmann},
      year={2019},
      eprint={1906.08172},
      archivePrefix={arXiv},
      primaryClass={cs.DC},
      url={https://arxiv.org/abs/1906.08172}, 
}

@misc{b3,
      title={MediaPipe Hands: On-device Real-time Hand Tracking}, 
      author={Fan Zhang and Valentin Bazarevsky and Andrey Vakunov and Andrei Tkachenka and George Sung and Chuo-Ling Chang and Matthias Grundmann},
      year={2020},
      eprint={2006.10214},
      archivePrefix={arXiv},
      primaryClass={cs.CV},
      url={https://arxiv.org/abs/2006.10214}, 
}

@misc{b4,
      title={You Only Look Once: Unified, Real-Time Object Detection}, 
      author={Joseph Redmon and Santosh Divvala and Ross Girshick and Ali Farhadi},
      year={2016},
      eprint={1506.02640},
      archivePrefix={arXiv},
      primaryClass={cs.CV},
      url={https://arxiv.org/abs/1506.02640}, 
}

@article{b5,
author = {Yang, Panle},
year = {2025},
month = {01},
pages = {762},
title = {Integrating intelligent algorithms in music education to analyze and improve posture and motion in instrumental training},
volume = {22},
journal = {Molecular \& Cellular Biomechanics},
doi = {10.62617/mcb762}
}

@article{b6,
author = {Rozé, Jocelyn and Aramaki, Mitsuko and Kronland-Martinet, Richard and Ystad, Sølvi},
year = {2020},
month = {08},
pages = {13882},
title = {Cellists’ sound quality is shaped by their primary postural behavior},
volume = {10},
journal = {Scientific Reports},
doi = {10.1038/s41598-020-70705-8}
}

@article{b7,
author = {Figueres, Jose and Perez-Soriano, Pedro and Belloch, Salvador and Figueres, Emili},
year = {2011},
month = {10},
pages = {23-34},
title = {Injuries Prevention in String Players},
volume = {4},
journal = {Journal of Sport and Health Research}
}

@misc{b8,
      title={EfficientNet: Rethinking Model Scaling for Convolutional Neural Networks}, 
      author={Mingxing Tan and Quoc V. Le},
      year={2020},
      eprint={1905.11946},
      archivePrefix={arXiv},
      primaryClass={cs.LG},
      url={https://arxiv.org/abs/1905.11946}, 
}

@article{b9,
      title={MobileNets: Efficient Convolutional Neural Networks for Mobile Vision Applications}, 
      author={Andrew G. Howard and Menglong Zhu and Bo Chen and Dmitry Kalenichenko and Weijun Wang and Tobias Weyand and Marco Andreetto and Hartwig Adam},
      year={2017},
      eprint={1704.04861},
      archivePrefix={arXiv},
      primaryClass={cs.CV},
      url={https://arxiv.org/abs/1704.04861}, 
}

@article{b10,
   title={Empowering Edge Intelligence: A Comprehensive Survey on On-Device AI Models},
   volume={57},
   ISSN={1557-7341},
   url={http://dx.doi.org/10.1145/3724420},
   DOI={10.1145/3724420},
   number={9},
   journal={ACM Computing Surveys},
   publisher={Association for Computing Machinery (ACM)},
   author={Wang, Xubin and Tang, Zhiqing and Guo, Jianxiong and Meng, Tianhui and Wang, Chenhao and Wang, Tian and Jia, Weijia},
   year={2025},
   month=apr, pages={1–39} 
}

@misc{b11,
      title={OpenPose: Realtime Multi-Person 2D Pose Estimation using Part Affinity Fields}, 
      author={Zhe Cao and Gines Hidalgo and Tomas Simon and Shih-En Wei and Yaser Sheikh},
      year={2019},
      eprint={1812.08008},
      archivePrefix={arXiv},
      primaryClass={cs.CV},
      url={https://arxiv.org/abs/1812.08008}, 
}

@article{b12,
  author={Johnson, David and Damian, Daniela and Tzanetakis, George},
  journal={Computer Music Journal}, 
  title={Detecting Hand Posture in Piano Playing Using Depth Data}, 
  year={2020},
  volume={43},
  number={1},
  pages={59-78},
  doi={10.1162/comj_a_00500}
}

@inproceedings{b13,
author = {Nielsen, Jakob and Molich, Rolf},
title = {Heuristic evaluation of user interfaces},
year = {1990},
isbn = {0201509326},
publisher = {Association for Computing Machinery},
address = {New York, NY, USA},
url = {https://doi.org/10.1145/97243.97281},
doi = {10.1145/97243.97281},
booktitle = {Proceedings of the SIGCHI Conference on Human Factors in Computing Systems},
pages = {249–256},
numpages = {8},
location = {Seattle, Washington, USA},
series = {CHI '90}
}

@misc{b14,
  author    = {Nielsen, Jakob},
  title     = {10 Usability Heuristics for User Interface Design},
  howpublished = {\url{https://www.nngroup.com/articles/ten-usability-heuristics/}},
  year      = {1994},
  note      = {Accessed: 2026-01-20}
}

@inproceedings{b15,
author = {Amershi, Saleema and Weld, Dan and Vorvoreanu, Mihaela and Fourney, Adam and Nushi, Besmira and Collisson, Penny and Suh, Jina and Iqbal, Shamsi and Bennett, Paul N. and Inkpen, Kori and Teevan, Jaime and Kikin-Gil, Ruth and Horvitz, Eric},
title = {Guidelines for Human-AI Interaction},
year = {2019},
isbn = {9781450359702},
publisher = {Association for Computing Machinery},
address = {New York, NY, USA},
url = {https://doi.org/10.1145/3290605.3300233},
doi = {10.1145/3290605.3300233},
booktitle = {Proceedings of the 2019 CHI Conference on Human Factors in Computing Systems},
pages = {1–13},
numpages = {13},
location = {Glasgow, Scotland Uk},
series = {CHI '19}
}

@misc{b16,
      title={SAM 3: Segment Anything with Concepts}, 
      author={Nicolas Carion and Laura Gustafson and Yuan-Ting Hu and Shoubhik Debnath and Ronghang Hu and Didac Suris and Chaitanya Ryali and Kalyan Vasudev Alwala and Haitham Khedr and Andrew Huang and Jie Lei and Tengyu Ma and Baishan Guo and Arpit Kalla and Markus Marks and Joseph Greer and Meng Wang and Peize Sun and Roman Rädle and Triantafyllos Afouras and Effrosyni Mavroudi and Katherine Xu and Tsung-Han Wu and Yu Zhou and Liliane Momeni and Rishi Hazra and Shuangrui Ding and Sagar Vaze and Francois Porcher and Feng Li and Siyuan Li and Aishwarya Kamath and Ho Kei Cheng and Piotr Dollár and Nikhila Ravi and Kate Saenko and Pengchuan Zhang and Christoph Feichtenhofer},
      year={2025},
      eprint={2511.16719},
      archivePrefix={arXiv},
      primaryClass={cs.CV},
      url={https://arxiv.org/abs/2511.16719}, 
}

@misc{b17,
  author    = {Wu, Y. and Kirillov, A. and Massa, F. and Lo, W.-Y. and Girshick, R.},
  title     = {Detectron2},
  howpublished = {\url{https://github.com/facebookresearch/detectron2}},
  year      = {2019},
  note      = {Computer software}
}

@misc{b18,
      title={You Only Look Once: Unified, Real-Time Object Detection}, 
      author={Joseph Redmon and Santosh Divvala and Ross Girshick and Ali Farhadi},
      year={2016},
      eprint={1506.02640},
      archivePrefix={arXiv},
      primaryClass={cs.CV},
      url={https://arxiv.org/abs/1506.02640}, 
}

@misc{b19,
      title={DETRs Beat YOLOs on Real-time Object Detection}, 
      author={Yian Zhao and Wenyu Lv and Shangliang Xu and Jinman Wei and Guanzhong Wang and Qingqing Dang and Yi Liu and Jie Chen},
      year={2024},
      eprint={2304.08069},
      archivePrefix={arXiv},
      primaryClass={cs.CV},
      url={https://arxiv.org/abs/2304.08069}, 
}

@misc{b20,
      title={What is YOLOv8: An In-Depth Exploration of the Internal Features of the Next-Generation Object Detector}, 
      author={Muhammad Yaseen},
      year={2024},
      eprint={2408.15857},
      archivePrefix={arXiv},
      primaryClass={cs.CV},
      url={https://arxiv.org/abs/2408.15857}, 
}

@misc{b21,
      title={YOLOv11: An Overview of the Key Architectural Enhancements}, 
      author={Rahima Khanam and Muhammad Hussain},
      year={2024},
      eprint={2410.17725},
      archivePrefix={arXiv},
      primaryClass={cs.CV},
      url={https://arxiv.org/abs/2410.17725}, 
}

@misc{b22,
      title={YOLOv12: Attention-Centric Real-Time Object Detectors}, 
      author={Yunjie Tian and Qixiang Ye and David Doermann},
      year={2025},
      eprint={2502.12524},
      archivePrefix={arXiv},
      primaryClass={cs.CV},
      url={https://arxiv.org/abs/2502.12524}, 
}

@article{b23,
  author    = {Bradski, G.},
  title     = {The OpenCV Library},
  journal   = {Dr. Dobb's Journal of Software Tools},
  year      = {2000}
}
\end{document}